\documentclass[aps,prl,showpacs,twocolumn,preprintnumbers,superscriptaddress,amssymb]{revtex4}
\usepackage{graphicx}
\usepackage{dcolumn}
\usepackage{bm}
\usepackage{amsmath}

\begin{document}

\title{Semiclassical backreaction around a nearly spinning cosmic string}
\author{Vitorio A. De Lorenci}
 \email{delorenci@unifei.edu.br}
\affiliation{Instituto de Ci\^encias Exatas, Universidade Federal de Itajub\'a, 
Av. BPS 1303 Pinheirinho, 37500-903 Itajub\'a, MG, Brazil}
\affiliation{PH Department, TH Unit, CERN, 1211 Geneva 23, Switzerland}
\author{Edisom S. Moreira, Jr.}
 \email{moreira@unifei.edu.br}
\affiliation{Instituto de Ci\^encias Exatas, Universidade Federal de Itajub\'a, 
Av. BPS 1303 Pinheirinho, 37500-903 Itajub\'a, MG, Brazil}
\affiliation{The Blackett Laboratory, Imperial College London, 
Prince Consort Road, London SW7 2AZ, U.K.}

\date{November, 2008}

\begin{abstract}

This paper investigates semiclassical backreaction 
of a conformally coupled massless scalar field on the 
geometrical background of a nearly spinning cosmic string ---
the spin density is smaller than, but arbitrarily close to, the
dislocation parameter. As the spin density approaches
the dislocation parameter, it is shown that an  ergoregion 
spreads indefinitely around the cosmic string, 
boosting along the string axis the once static observers. Considering that
the geometrical background contains closed timelike curves
when the spin density exceeds the dislocation parameter,
it is argued that the appearance of the ergoregion 
may be part of a chronology protection mechanism
that takes place in related non stationary geometries.

\end{abstract}
\pacs{04.62.+v, 04.20.Gz, 11.27.+d}
\maketitle

\paragraph{{\bf I - Introduction}.}
%
Cosmic strings are objects which may play relevant role in astrophysics, 
cosmology, and fundamental physics \cite{vil00,kib05}.
It has long been noticed that such objects offer a rich arena to investigate
the interplay between non trivial global geometry and quantum field theory \cite{dow77,his87}.
Since gravitational fields generated by cosmic strings correspond to locally flat
backgrounds (geometrical analogs of the Aharonov-Bohm setup), 
calculations usually turn out to be simpler than those in locally curved spacetime,
leading to quantum effects due to a nonvanishing global curvature.
Recent investigations on quantum fields around cosmic strings have addressed 
massive fields, higher spins, various dimensions and boundary conditions, among other issues
(see, e.g., \cite{mel06}).

As is well known, the geometry of spacetime outside an ordinary cosmic string is given by the line element \cite{vil00,his85},
\begin{equation}
ds^{2}=d\tau^{2}-dr^{2}-\alpha^{2}r^{2}d\theta^{2}-d\xi^{2},
\label{ocone}
\end{equation}
where the disclination parameter $\alpha$ is related to the mass density $\mu$ of the straight string 
by $\alpha=1-4\mu$ 
(units as in \cite{his87} will be used, i.e., $G=c=1$).
The coordinates in Eq. (\ref{ocone}) have the same nature as those in the Minkowski line element 
(when expressed in terms of cylindrical coordinates), with an important difference that Eq. (\ref{ocone})
hides a conical singularity at $r=0$, corresponding to a deficit angle $2\pi(1-\alpha)$, if $\mu\neq 0$.
According to the physics of formation of ordinary cosmic strings, $\alpha$ is very close to one \cite{vil00,kib05}.
It should be remarked that words such as ``disclination'' and ``dislocation'' 
have been borrowed from condensed matter physics, where geometrical aspects also appear 
(see, e.g., \cite{kro81}).

The metric tensor in Eq. (\ref{ocone}) is cylindrically symmetric
and invariant under boosts along the symmetry axis \cite{vil00,his85}. 
Definition of a new angle as,
\begin{equation}
\varphi:=\alpha\theta\hspace{1.7cm} \varphi \sim \varphi+2\pi\alpha,
\label{nangle}
\end{equation}
clearly shows that  Eq. (\ref{ocone}) corresponds to a locally flat vacuum solution of 
the Einstein equations. 
If the requirement of boost invariance is relaxed, 
one is led to a generalization of Eq. (\ref{ocone}) \cite{maz86,gal93,tod94},
\begin{equation}
ds^{2}=(d\tau+Sd\theta)^{2}-dr^{2}-\alpha^{2}r^{2}d\theta^{2}
-(d\xi+\kappa d\theta)^{2},
\label{ssframe}
\end{equation}
containing two new parameters (which will be taken to be non negative):
the spin density $S$, and the dislocation parameter $\kappa$.
When $S>\kappa$, Eq. (\ref{ssframe}) 
shows that the associated spacetime contains closed timelike curves (CTCs),
and therefore violates causality
[taking $d\tau=dr=d\xi=0$ in Eq. (\ref{ssframe}), 
CTCs are obtained if $r<\sqrt{S^{2}-\kappa^{2}}/\alpha$].
The locally flat character of Eq. (\ref{ssframe}) is revealed by considering
Eq. (\ref{nangle}), ${\rm T}:= \tau + S\theta$ and 
$\Xi:=\xi+ \kappa\theta$, resulting
\begin{equation}
ds^{2}=d{\rm T}^{2}-dr^{2}-r^{2}d\varphi^{2}-d\Xi^{2}.
\label{siframe}
\end{equation}

Setting $\kappa=0$ in Eq. (\ref{ssframe}) yields the geometry around  a
``spinning cosmic string'' \cite{maz86}, whose terminology has to do with the 
fact that, by omitting the last term in Eq. (\ref{ssframe}), the resulting
line element corresponds to the geometry around a particle 
with spin $S$ in $(2+1)$-dimensions \cite{des84}
(clearly both cases present CTCs). Setting instead
$S=0$, Eq. (\ref{ssframe}) becomes the line element corresponding to a 
``cosmic dislocation'' \cite{gal93}.

In fact, when $S\neq\kappa$, Eq. (\ref{ssframe}) describes 
either a cosmic dislocation, or a spinning cosmic string. 
For $S<\kappa$, 
the following Lorentz transformation in the $\tau-\xi$ plane,
\begin{eqnarray}
&&t=\frac{\tau-v\xi}{\sqrt{1-v^{2}}},\hspace{0.7cm}
z=\frac{\xi-v\tau}{\sqrt{1-v^{2}}},\hspace{0.7cm}v:=S/\kappa,
\label{boost}
\end{eqnarray} 
leads to
\begin{equation}
ds^{2}=dt^{2}-dr^{2}-\alpha^{2}r^{2}d\theta^{2}
-(dz+\kappa' d\theta)^{2},
\label{dle}
\end{equation}
describing the geometry of a cosmic dislocation
with dislocation parameter 
\begin{equation}
\kappa':=\sqrt{\kappa^{2}-S^{2}}.
\label{dparameter}
\end{equation}
If, on the other hand, $S>\kappa$, replacing $v$ in Eq. (\ref{boost}) 
by $\kappa/S$, Eq. (\ref{ssframe}) is recast as 
\begin{equation}
ds^{2}=(dt+S'd\theta)^{2}-dr^{2}-\alpha^{2}r^{2}d\theta^{2}-dz^{2},
\label{sle}
\end{equation}
corresponding to a spinning cosmic string with spin density 
$S':=\sqrt{S^{2}-\kappa^{2}}$.

The fact that vacuum fluctuations typically diverge 
when CTCs are about to form 
(for a review see \cite{mat02})
has led to the chronology protection conjecture \cite{haw92}, 
according to which the laws of physics do not allow the appearance
of ``time machines'' (if vacuum fluctuations are strong, backreaction effects
could modify the original geometry preventing the formation of CTCs).
Although Eq. (\ref{ssframe}) describes a stationary geometry,
the parameters can be adjusted in order to simulate 
a scenario where CTCs were about to form,
namely, considering $S<\kappa$ and taking $S\rightarrow\kappa$.
This approach has been used in \cite{lor04}, 
showing that the vacuum expectation value of the energy momentum tensor
of a massless scalar field diverges in the coordinate systems of 
Eqs. (\ref{ssframe}) and (\ref{siframe}), when $S\rightarrow\kappa$. 
However, it remains finite when expressed in terms of the coordinates in
Eq. (\ref{dle}) (this is expected since, when $\kappa'\rightarrow 0$, Eq. (\ref{dle}) 
approaches the line element of an ordinary cosmic string, for which 
vacuum fluctuations behave well \cite{dow77}) \cite{lor03}. 
It might appear that the divergent effect in the coordinate system of Eq. (\ref{ssframe}) 
is purely due to some relativistic factor coming from Eq. (\ref{boost}); but that is not the case.
The expressions for vacuum fluctuations in the background of a cosmic dislocation
carry a certain function which presents a mild divergence when its argument vanishes.
The transformation from the coordinates in Eq. (\ref{dle}) to those in Eq. (\ref{ssframe}) 
activates this divergence. As has been shown in \cite{lor04},
if the mentioned function were not divergent for a vanishing argument, 
as $S$ approached $\kappa$ the vacuum expectation value of the energy momentum tensor
in the coordinate system of Eq. (\ref{ssframe}) would remain finite, 
suggesting violation of chronology protection.

At first sight the study of the ``strong'' backreaction 
effects on the metric tensor in Eq. (\ref{ssframe}) seems to be intractable
(the procedure possibly becomes not reliable
by refeeding Einstein's equations with ``strong'' vacuum fluctuations).
However, taking into account the fact 
that backreaction on the metric tensor in Eq. (\ref{dle}) is ``weak'' \cite{his87,lor05}, 
one could first solve the problem in the coordinate system of Eq. (\ref{dle}),
then translating the results to that of  Eq. (\ref{ssframe}), 
via Eq. (\ref{boost}). This approach will be implemented in the following sections.

It should be added that vacuum fluctuations in the geometry of a
spinning cosmic string \cite{mat90,lor01} [cf. Eq. (\ref{sle})] 
are pathological due to the presence of CTCs
(the corresponding spacetime is nonglobally hyperbolic, 
and the usual quantization procedures lead to
divergent vacuum fluctuations in all frames \cite{lor01}).
CTCs also spoil unitarity of quantum mechanics in the corresponding
(2+1)-dimensional geometry \cite{ger89}.

In the next section, the study of semiclassical 
backreaction around a cosmic dislocation  \cite{lor05} 
is extended, and used in the following section to compute
quantum corrections in the metric tensor in Eq. (\ref{ssframe}),
when $S$ approaches $\kappa$ from below (``nearly spinning cosmic string'').
The effects on static observers are determined,
showing the appearance of a region around the cosmic string, 
whose features resemble those of the ergosphere of a rotating black hole. 
A summary and further discussion are presented in the last section.

\paragraph{{\bf II - Backreaction around cosmic dislocations}.}
\label{sec: cd}
In order to study  semiclassical backreaction
of a conformally coupled massless scalar field 
$\phi$ on the geometry of 
a cosmic dislocation \cite{lor05},
it is convenient to consider in Eq. (\ref{dle}) new coordinates, 
$Z := z + \kappa' \theta$ and Eq. (\ref{nangle}),
\begin{equation}
ds^{2}=dt^{2}-dr^2 -r^2 d\varphi^2 - dZ^2,
\label{fle}
\end{equation}
with $(t,r,\varphi,Z) \sim (t,r,\varphi+2\pi\alpha,Z+2\pi\kappa')$.
In terms of the local inertial coordinate system in Eq. (\ref{fle}),
the general form of the vacuum expectation value of the energy momentum tensor
for $\phi$ is given by \cite{lor03},
\begin{equation}
\left< {T}^{\mu}{}_{\nu} \right>(r) =
\left(
     \begin{array}{cccc}
\left<T^{t}{}_{t}\right>& 0 & 0 & 0\\
0 &\left<T^{r}{}_{r}\right> & 0  & 0 \\
0& 0 & \left<T^{\varphi}{}_{\varphi}\right> & \left<T^{\varphi}{}_{Z}\right> \\
0&  0 & r^{2}\left<T^{\varphi}{}_{Z}\right>& 
\left<T^{Z}{}_{Z}\right>
     \end{array}
\right),
\label{matrix}
\end{equation}
with two components related as 
\begin{equation}
\left<T^{Z}{}_{Z}\right>=\left<T^{t}{}_{t}\right>+\frac{\kappa'^2}{r^6}
\hbar f_{\alpha}\left(\kappa'^2/r^2\right),
\label{drelation}
\end{equation}
where $2\pi\hbar$ is the Planck constant, and
\begin{widetext}
\begin{equation}
f_\alpha(x):=
-\frac{1}{2}\int^{\infty}_{0}d\lambda\sum_{n=1}^{\infty}
\frac{n^2\left[\lambda^2-\pi^2(4\alpha^2 n^2-1)\right]}
{\left[\pi^2(2\alpha n+ 1)^2
+ \lambda^2\right] \left[\pi^2(2\alpha n - 1)^2 +\lambda^2\right]
\left[\cosh^2(\lambda/2) + n^2\pi^2 x\right]^3}.
\label{function}
\end{equation}
\end{widetext}
As $x\rightarrow 0$, $f_\alpha(x)\rightarrow +\infty$; but the divergence
is mild since $xf_\alpha(x)\rightarrow 0$ in this limit \cite{lor04}.

When the cosmic dislocation is absent, i.e., $\alpha=1$ and $\kappa'=0$,
$\left< T^{\mu}{}_{\nu} \right>$ vanishes.
If $\kappa'/r\ll 1$, one has approximately,
\begin{equation}
\left< T^{\mu}{}_{\nu} \right>(r) =\frac{\hbar}{r^4}
\left(
     \begin{array}{cccc}
      -A &  0 & 0                    & 0 \\
       0 & -A & 0                    & 0 \\
       0 &  0 & 3A                   & \kappa' B/r^2 \\
       0 &  0 & \kappa' B             & -A
     \end{array}
\right),
\label{emtensor}
\end{equation}
where $A(\alpha):=(\alpha^{-4}-1)/1440\pi^2$ 
and $B(\alpha)$ is defined as in Eq. (20)
of \cite{lor03} [$B(\alpha=1)=1/60\pi^2$]. 
When $\kappa'=0$, Eq. (\ref{emtensor}) reduces to the form
long known in the literature corresponding to an ordinary cosmic string \cite{dow77}.
Subleading contributions in Eq. (\ref{emtensor}) must be considered
if $\alpha=1$ and $\kappa'\neq 0$.

It should be remarked that the study of 
vacuum polarization around a cosmic dislocation
has been implemented using the vacuum associated with
the time coordinate $t$ in Eq. (\ref{dle}). 
(In fact, by observing Eqs. (\ref{ssframe}) and (\ref{dle}), 
one sees that the generators of translations in $\tau$ and in $t$
are globally timelike Killing vector fields that commute, and therefore
defining the same vacuum state \cite{chm94}.)
This appears to be a natural choice of vacuum, since when $\kappa'=0$
and $\alpha=1$, the corresponding Feynman propagator becomes the Minkowski
propagator \cite{lor03}.

Observing the most general form of a 
static and cylindrically symmetric line element \cite{ste03},
one now allows quantum perturbations $\gamma_{\mu\nu}(r)$ 
(which are assumed to be linear in $\hbar$) of the 
background metric tensor in Eq. (\ref{fle}),
\begin{eqnarray}
ds^{2}&=&(1+\gamma_{tt})dt^{2}+(-1+\gamma_{rr})dr^{2}
-r^{2}d\varphi^{2}
\nonumber \\ &&
+2\gamma_{Z \varphi} dZ d\varphi
+(-1+\gamma_{ZZ})dZ^{2},
\label{qcorrections}
\end{eqnarray}
involving four unknown functions of $r$.
Einstein's equations $R^{\mu}{}_{\nu}=-8\pi \left< T^{\mu}{}_{\nu} \right>$ are then 
fed with traceless (since spacetime is locally flat) 
$\left< {T}^{\mu}{}_{\nu} \right>$ in Eq. (\ref{matrix}),
leading to the following set of linearized Einstein's equations
\begin{eqnarray}
&r^{2}\gamma_{tt,rr}+r\gamma_{tt,r}=16\pi r^{2}\left<T^{t}{}_{t}\right>,&
\label{einstein1}
\\
&r\gamma_{rr,r}+r\gamma_{tt,r}-r\gamma_{ZZ,r}=16\pi r^{2}\left<T^{\varphi}{}_{\varphi}\right>,&
\label{einstein2}
\\
&r^{2}\gamma_{\varphi Z,rr}-r\gamma_{\varphi Z,r}=-16\pi r^{4}\left<T^{\varphi}{}_{Z}\right>,&
\label{einstein3}
\\
&r^{2}\gamma_{ZZ,rr}+r\gamma_{ZZ,r}=-16\pi r^{2}\left<T^{Z}{}_{Z}\right>.&
\label{einstein4}
\end{eqnarray}
The equation involving $\left<T^{r}{}_{r}\right>$ has been omitted, since it follows from
Eqs. (\ref{einstein1}), (\ref{einstein2}) and (\ref{einstein4}), and by considering that 
$\left<{T}^{\mu}{}_{\nu} \right>$ is convariantly conserved. 

Eqs. (\ref{einstein1}), (\ref{einstein3}) and (\ref{einstein4})
have the form of Euler's equation
$x^{2}y''+axy'=G(x)$, whose general solution can be written as
\begin{equation}
y(x)=c_{1}+c_{2}x^{1-a}+\frac{1}{a-1}\int_{\beta}^{x}\frac{G(u)}{u}\left[1-\left( \frac{x}{u}\right)^{1-a}\right]du
\label{eulera}
\end{equation}
if $a\neq 1$, or
\begin{equation}
y(x)=c_{1}+c_{2}\log x+\int_{\beta}^{x}\frac{G(u)}{u}\log\left(\frac{x}{u}\right) du
\label{eulerb}
\end{equation}
if $a=1$. Eqs. (\ref{eulera}) and (\ref{eulerb}) contain arbitrary constants $c_{1}$ and $c_{2}$, and a point $\beta$
that can be conveniently chosen.

The solutions of Eqs. (\ref{einstein1}) and (\ref{einstein4}) can be read from Eq. (\ref{eulerb}),
\begin{eqnarray}
\gamma_{tt}&=&16\pi\int_{r}^{\infty}u\left<T^{t}{}_{t}\right>\log\left(\frac{u}{r}\right) du,
\label{seinstein14a}
\\ 
\gamma_{ZZ}&=&-16\pi\int_{r}^{\infty}u\left<T^{Z}{}_{Z}\right>\log\left(\frac{u}{r}\right) du,
\label{seinstein14}
\end{eqnarray}
where the constants have been found to vanish by applying
a dimensional argument used in the 
backreaction problem around an ordinary cosmic string \cite{his87}.
To illustrate the procedure, one assumes $c_{2}=0$ but  $c_{1}\neq 0$. Now, $c_{1}$
is dimensionless and linear in $\hbar$,
and the only dimensionful parameters in Eqs. (\ref{einstein1})-(\ref{einstein4}) are $\hbar$ and $\kappa'$
(with units of squared length and length, respectively). It follows that 
$c_{1}=c_{o} \hbar/\kappa'^{2}$, where 
$c_{o}$ is dimensionless. Clearly that is not acceptable unless $c_{o}$ vanishes:
by setting $\kappa'=0$, the (finite) results corresponding to an ordinary cosmic string should be reproduced.

Similar considerations regarding Eqs.  (\ref{einstein3}) and (\ref{eulera}) lead to 
\begin{eqnarray}
&&\gamma_{\varphi Z}=8\pi\int_{r}^{\infty}u\left<T^{\varphi}{}_{Z}\right>(r^{2}-u^{2})du,
\label{seinstein3}
\end{eqnarray}
and combination of Eqs. (\ref{seinstein14a})-(\ref{seinstein14})  yields
\begin{equation}
\label{seinstein2}
\gamma_{rr}=16\pi\int_{\infty}^{r}u
\left[\left(\left<T^{t}{}_{t}\right> + \left<T^{Z}{}_{Z}\right>\right)\log\left(\frac{u}{r}\right)
+\left<T^{\varphi}{}_{\varphi}\right>\right]du
\end{equation}
by solving Eq. (\ref{einstein2}).
It is worth noting that the argument of $\left<T^{\mu}{}_{\nu}\right>$ 
in Eqs. (\ref{seinstein14a})-(\ref{seinstein2}) is the integration parameter $u$, i.e.,
$\left<T^{\mu}{}_{\nu}\right>(u)$.

Observing Eq. (\ref{drelation}), the following relation arises from Eqs. (\ref{seinstein14a})-(\ref{seinstein14})
\begin{equation}\gamma_{tt}+\gamma_{ZZ}=-\kappa'^{2}\hbar F_{\alpha}(\kappa',r),
\label{tzrelation}
\end{equation}
where the function
\begin{equation}
F_{\alpha}(\kappa',r):=16\pi\int_{r}^{\infty}\frac{1}{u^{5}}f_{\alpha}\left(\kappa'^{2}/u^{2}\right)\log\left(\frac{u}{r}\right)du
\label{F}
\end{equation}
diverges positively when $\kappa'\rightarrow 0$. However,
in the limit when $\kappa'\rightarrow 0$,  $\gamma_{tt}=-\gamma_{ZZ}$.

One can check the consistency of these results by taking into 
Eqs. (\ref{seinstein14a})-(\ref{seinstein2}) 
the expressions for $\left<T^{\mu}{}_{\nu}\right>(r)$ in Eq. (\ref{emtensor}).
Performing the integrations, it follows that
\begin{eqnarray}
ds^2 &=& \left(1-\frac{4\pi A\hbar }{r^2}\right)(dt^2-dZ^2) 
-\left(1+\frac{16\pi A\hbar}{r^2}\right)dr^2
\nonumber \\
&&-r^2 d\varphi^2 -\frac{4\pi\kappa' B \hbar}
{r^2} \ d\varphi\ dZ,
\label{breaction1}
\end{eqnarray}
reproducing the results in \cite{his87,lor05}.

At this point, it should be stressed  that in order the semiclassical scheme,
based on the use of the linearized Einstein equations, to make sense
the ``perturbations'' in Eq. (\ref{qcorrections}) must be tiny 
(i.e.,  $\gamma_{tt}\ll1$, $\gamma_{rr}\ll1$, $\gamma_{ZZ}\ll1$, and $\gamma_{Z\varphi}\ll\kappa'$ for nonvanishing $\kappa'$).
Thus, examining Eq. (\ref{breaction1}), $\hbar/r^{2}\ll1$ is assumed to hold outside the cosmic dislocation.

\paragraph{\- {\bf III - Backreaction around nearly spinning cosmic strings}.}
\label{sec: ergo}
It follows from Eq. (\ref{boost}) that the inertial coordinate systems in Eqs. (\ref{siframe}) and (\ref{fle})
are related by
\begin{eqnarray}
&&t=\frac{{\rm T}-v\Xi}{\sqrt{1-v^{2}}},\hspace{0.5cm}
Z=\frac{\Xi-v{\rm T}}{\sqrt{1-v^{2}}},\hspace{0.5cm}v:=S/\kappa.
\label{iboost}
\end{eqnarray} 
Thus the line element in Eq. (\ref{qcorrections}) can be recast as
\begin{eqnarray}
ds^{2}=(1+\gamma_{tt}-S^{2}\hbar F_{\alpha})d{\rm T}^{2}
+(-1+\gamma_{rr})dr^{2}
\nonumber\\ 
- r^{2}d\varphi^{2}
-2(S/\kappa')\gamma_{Z\varphi}d{\rm T} d\varphi+2S\kappa\hbar F_{\alpha}d{\rm T}d\Xi
\nonumber \\
+ 2(\kappa/\kappa')\gamma_{Z\varphi}d\Xi d\varphi
+(-1+\gamma_{ZZ}-S^{2}\hbar F_{\alpha})d\Xi^{2},
\label{Qcorrections}
\end{eqnarray}
where Eqs. (\ref{dparameter}) and (\ref{tzrelation}) have been used.
For given values of the parameters $\kappa$ and $S$, when $r\rightarrow\infty$ the line element 
in Eq. (\ref{Qcorrections}) reduces to the flat form in Eq. (\ref{siframe}), as can be seen 
from the expressions for $\gamma_{\mu\nu}$ and $F_{\alpha}$ in the previous section.

To obtain quantum corrections in the
coordinate system 
$\{\tau,r,\theta,\xi\}$
of Eq. (\ref{ssframe}), one simply
replaces in Eq. (\ref{Qcorrections}) $d{\rm T}$,  $d\varphi$ and   $d\Xi$ by
$d\tau + Sd\theta$, $\alpha d\theta$ and 
$d\xi+ \kappa d\theta$, respectively [see text just before Eq. (\ref{siframe})].
An observer that moves at most axially (i.e., with $dr=0$ and $d\theta=0$) has $ds^{2}>0$ given by
\begin{eqnarray}
ds^{2}&=&(1+\gamma_{tt}-S^{2}\hbar F_{\alpha})d \tau^{2}
+2S\kappa\hbar F_{\alpha}d\tau d\xi
\nonumber \\
&&+(-1+\gamma_{ZZ}-S^{2}\hbar F_{\alpha})d\xi^{2}.
\label{ptime}
\end{eqnarray}
If the observer is at rest, it follows from Eq. (\ref{ptime}) that 
$ds^{2}=(1+\gamma_{tt}-S^{2}\hbar F_{\alpha})d \tau^{2}$. 
By letting $S$ grow toward $\kappa\neq 0$, the latter kept fixed, and 
recalling that $\gamma_{tt}\rightarrow -4\pi A\hbar/r^{2}$
and $F_{\alpha}\rightarrow +\infty$ when $\kappa'\rightarrow 0$
[see Eqs. (\ref{dparameter}), (\ref{F}) and (\ref{breaction1})], 
it becomes clear that there is a value of $S$
above which $ds^{2}$ becomes negative, and therefore the observer
cannot remain at rest (static). In the region defined by
\begin{equation}
 1+\gamma_{tt}(r)-S^{2}\hbar F_{\alpha}(\kappa',r)<0
\label{ergos}
\end{equation}
(and at its surface) there is no static observers,
resembling in this sense 
the ergosphere of a rotating black hole (see, e.g., \cite{whe73}) ---
the time translation Killing vector field $\chi^{\mu}=(1,0,0,0)$ is not timelike, i.e.,
\begin{equation}
\chi^{2}:=\chi^{\mu}\chi_{\mu}= 1+\gamma_{tt}-S^{2}\hbar F_{\alpha}<0.
\label{killing}
\end{equation}
Since  $F_{\alpha}(\kappa',r)$ is a decreasing function of $r$,
as $S$ approaches $\kappa$ (i.e., $\kappa'\rightarrow 0$)
the ergoregion defined in Eq. (\ref{ergos}) widens indefinitely throughout the space.

In order to study further the properties of the ergoregion, 
the axial velocity of an observer,
\begin{equation}
V:=\frac{d\xi}{d\tau},
\label{avelocity}
\end{equation}
is used to rewrite  Eq. (\ref{ptime}) as
$ds^{2}=Pd\tau^{2}$, where
\begin{eqnarray}
P(V)&:=&1+\gamma_{tt}-S^{2}\hbar F_{\alpha}
+2S\kappa\hbar F_{\alpha}V
\nonumber \\
&&+(-1+\gamma_{ZZ}-S^{2}\hbar F_{\alpha})V^{2}.
\label{p}
\end{eqnarray}
As $\gamma_{ZZ}\ll 1$, the coefficient 
$-1+\gamma_{ZZ}-S^{2}\hbar F_{\alpha}$ is negative, resulting that
$V$ must be between the roots of $P(V)$, such that $ds^{2}>0$:
\begin{equation}
V_{-}<V<V_{+}
\label{roots0} 
\end{equation}
with
\begin{equation}
V_{\pm}=\frac{S\kappa\hbar F_{\alpha}\pm\sqrt{(1-\gamma_{ZZ})(1+\gamma_{tt})}}
{1-\gamma_{ZZ}+S^{2}\hbar F_{\alpha}}.
\label{roots}
\end{equation}
In deriving $V_{\pm}$ in Eq. (\ref{roots}), Eq. (\ref{tzrelation}) has been used.

Observing the expressions in the previous section, one sees from Eq. (\ref{roots})
that far away from the cosmic string (i.e., when $r\rightarrow\infty$) 
$V_{\pm}\rightarrow\pm 1$,
the usual Minkowski limits in both directions. Inside the ergoregion, 
it follows from Eqs. (\ref{dparameter}), (\ref{tzrelation}) and (\ref{ergos}) that
$1-\gamma_{ZZ}-k^{2}\hbar F_{\alpha}<0$. This inequality combined with that
in Eq. (\ref{ergos}) leads to $V_{-}>0$, i.e., both $V_{+}$ and $V_{-}$ are positive --- 
in the ergoregion the observer must be moving in the positive direction.
By letting $S\rightarrow\kappa$ in Eq. (\ref{roots}), it results that
$V_{-}$  and $V_{+}$ tend to merge to unity:
$V_{\pm}\rightarrow 1$ as $\kappa'\rightarrow 0$.

Other properties of the ergoregion are revealed by considering the
energy 
$E=mg_{\tau{}\mu}dx^{\mu}/ds$
of a particle of mass $m$ 
(see, e.g., \cite{lan75})
constrained to move axially, and thus with proper time
given by Eq. (\ref{ptime}),
\begin{equation}
E(V)=\frac{m}{\sqrt{P(V)}}
\left(1+\gamma_{tt}-S^{2}\hbar F_{\alpha}
+S\kappa\hbar F_{\alpha}V\right).
\label{energy}
\end{equation}
If the particle travels with speed
\begin{equation}
V_{o}:=\frac{V_{+}+V_{-}}{2}=
\frac{S\kappa\hbar F_{\alpha}}
{1-\gamma_{ZZ}+S^{2}\hbar F_{\alpha}},
\label{mvelocity}
\end{equation}
it results that
\begin{equation}
E(V_{o})=m\sqrt{P(V_{o})},
\label{renergy}
\end{equation}
where Eq. (\ref{p}) has been used.
Noting Eq (\ref{tzrelation}), one obtains
\begin{equation}
P(V_{o})=\frac{(1-\gamma_{ZZ})(1+\gamma_{tt})}
{1-\gamma_{ZZ}+S^{2}\hbar F_{\alpha}}.
\label{pvzero}
\end{equation}
When $r\rightarrow\infty$, it follows 
from Eqs. (\ref{mvelocity})-(\ref{pvzero}) 
that $E\rightarrow m$, 
the usual particle rest energy corresponding to $V_{o}\rightarrow 0$.
On the other hand, when $r$ is such that the particle
is in the ergoregion, $S\rightarrow \kappa$ yields
$E(V_{o})\rightarrow 0$
and $V_{o}\rightarrow 1$. 

One can also derive from Eq. (\ref{energy}) that outside the ergoregion,
where $V_{-}<0$, $E(V\rightarrow V_{\pm})\rightarrow +\infty$.
And inside the ergoregion, where $V_{\pm}>0$,
$E(V\rightarrow V_{\pm})\rightarrow \pm\infty$,
vanishing when $V=(-1-\gamma_{tt}+S^{2}\hbar F_{\alpha})/S\kappa\hbar F_{\alpha}$.

Although this section addresses pure axial motion only, 
a rather straightforward calculation shows that an observer cannot have
pure radial motion in the ergoregion. And, if the observer has initially
pure circular motion in the ergoregion, it will eventually become helical
as $S\rightarrow\kappa$.

\paragraph{{\bf IV - Conclusion}.}
\label{sec: disc}
In this work, semiclassical backreaction 
on the metric tensor in Eq. (\ref{ssframe})
was determined, by extending a previous calculation on the geometrical 
background of a cosmic dislocation, Eq. (\ref{dle}). When $S>\kappa$ in Eq. (\ref{ssframe}),
the corresponding spacetime is nonglobally hyperbolic, since it contains CTCs.
It was shown that when $S$ approaches $\kappa$ from below, 
due to backreaction, a  cylindrical ergoregion
spreads around the ``nearly spinning cosmic string'', eventually covering the whole space.
This is encapsulated in the rather unexpected fact that $\chi^{2}=1$ before backreaction
is taken into account, whereas after backreaction is taken  into account [cf. Eq. (\ref{killing})],
\begin{equation}
\chi^{2}\rightarrow -\infty,
\label{result}
\end{equation}
as $S\rightarrow \kappa$ (for a fixed $r$).

In the coordinate system of  Eq. (\ref{ssframe}),
Eq. (\ref{result}) 
[which obviously is a coordinate independent statement]
is interpreted as strong backreaction effects, on 
the background metric tensor, resulting
from amplifications of the weak  quantum corrections
$\gamma_{\mu\nu}$ in Eq. (\ref{qcorrections}).
Expressions such as Eqs. 
(\ref{drelation}) and (\ref{tzrelation})
are not affected by the divergence of 
$f_{\alpha}(x)$  [cf. Eq. (\ref{function})]
due to the factor $\kappa'^{2}$
(since the divergence is weaker than $1/x$).
However, the transformation (\ref{boost}) 
``replaces'' $\kappa'^{2}$ by $S^{2}$, $\kappa^{2}$ or $\kappa S$, 
exposing the divergence which
causes the strong backreaction effects
on the metric tensor in Eq. (\ref{ssframe}),
when $S\rightarrow\kappa$. 
This divergence is also responsible
for the divergent $\left< {\cal T}^{\mu}{}_{\nu} \right>$, 
as has been shown in \cite{lor04}.

Before backreaction is considered, 
by  letting $S$ in Eq. (\ref{ssframe}) grow toward $\kappa$,
and eventually becoming greater than $\kappa$, 
it follows that a static observer
would see the transition between the two non equivalent geometries 
in Eqs. (\ref{dle}) and  (\ref{sle}),
simulating the appearance of a ``time machine''.
After backreaction is considered in the geometry of 
Eq. (\ref{ssframe}), the picture changes radically.
For a fixed $r$, Eq. (\ref{ptime}) shows
that the metric tensor diverges as $S\rightarrow\kappa$.
Moreover, static observers are only possible outside the ergoregion,
which as $S\rightarrow\kappa$
widens indefinitely across the space, dragging 
along the cosmic string, in the positive direction,
the once static observers.
This new picture seems to suggest that no observer would detect
the appearance of a ``time machine''. In other words,
in related non stationary geometries,  the ergoregion
and its associated strong effects 
would be part of a chronology protection mechanism.

Some remarks regarding the coordinate systems 
in Eqs. (\ref{ssframe}) and (\ref{dle}) are in order. 
Recalling that the Killing vector field $\chi^{\mu}$
is the generator of translations in the time $\tau$
(and not in the time $t$), it should be clear that
the dragging of static observers by the ergoregion takes place 
only in the coordinate system $\{\tau,r,\theta,\xi\}$ 
[see Eq. (\ref{ptime})]. In the coordinate system of Eq. (\ref{dle}),
backreaction effects when $S\rightarrow\kappa$ are those
(tiny effects) around an ordinary cosmic string [obtained by setting
$\kappa'=0$ and $d\varphi=\alpha d\theta$ in Eq. (\ref{breaction1})],
resulting that the generator of translations in $t$ remains globally timelike,
and observers once at rest can stay at rest.
Note, however, that any coordinate system
does have its own interpretation of the ergoregion, namely,
an observer initially following an integral curve of
$\chi^{\mu}$ will depart from it where $\chi^{2}\leq 0$.

It should be stressed that
the static geometry of Eq. (\ref{dle})
is not equivalent to the stationary (but not static)
geometry of Eq. (\ref{sle}). 
If one wishes to simulate
a non stationary scenario where a transition 
possibly takes place from Eq. (\ref{dle}) to Eq. (\ref{sle}), then
the natural setting to do so is Eq. (\ref{ssframe})
which describes either geometries.

\begin{acknowledgments}
VADL and ESM are grateful, respectively, to Sergio Ferrara (Theory Division at CERN)
and to Chris Hull (Theory Group at Imperial College London) for their hospitality.
This work was partially supported by the research agencies CNPq, CAPES and FAPEMIG.
\end{acknowledgments}


\begin{thebibliography}{17}


\bibitem{vil00}
A. Vilenkin and E. P. S. Shellard,
{\it Cosmic Strings and Other Topological Defects}
(Cambridge University Press, Cambridge, England, 1994).

\bibitem{kib05}
T. W. B. Kibble, arXiv:astro-ph/0410073; A. Vilenkin, arXiv:hep-th/0508135.

\bibitem{dow77}
J. S. Dowker, J. Phys. A \textbf{10}, 115 (1977);
T. M. Helliwell and D. A. Konkowski,
Phys. Rev. D \textbf{34}, 1918 (1986);
B. Linet, {\it ibid}. \textbf{35}, 536 (1987);
V. P. Frolov and E. M. Serebriany, 
 {\it ibid}. \textbf{35}, 3779 (1987);
J. S. Dowker,
{\it ibid}. \textbf{36}, 3095 (1987); {\it ibid}. \textbf{36}, 3742 (1987);
A. G. Smith, in 
{\it The Formation and Evolution of Cosmic Strings},
Proceedings of the Cambridge Workshop, Cambridge, England,
edited by G. W. Gibbons, S. W. Hawking, and T. Vachaspati
(Cambridge University Press, Cambridge, England, 1990).

\bibitem{his87}
W. A. Hiscock, Phys. Lett. B {\bf 188}, 317 (1987).

\bibitem{mel06}
V. B. Bezerra and N. R. Khusnutdinov, Class. Quantum Grav.  \textbf{23}, 3449 (2006);
E. R. Bezerra de Mello and A. A. Saharian, Phys. Lett. B \textbf{642}, 129 (2006);
E. R. Bezerra de Mello, V. B. Bezerra, A. A. Saharian, and A. S. Tarloyan, Phys. Rev. D \textbf{74}, 025017 (2006);
E. R. Bezerra de Mello, V. B. Bezerra,  and A. A. Saharian, Phys. Lett. B \textbf{645}, 245 (2007);
J. Spinelly and E. R. Bezerra de Mello, JHEP \textbf{09}, 005 (2008);
E. R. Bezerra de Mello and A. A. Saharian, Phys. Rev. D \textbf{78}, 045021 (2008);
E. R. Be\-zer\-ra de Mello, V. B. Bezerra, A. A. Saharian, and A. S. Tarloyan, {\it ibid}. \textbf{78}, 105007 (2008).
\bibitem{his85}
W. A. Hiscock, Phys. Rev. D {\bf 31}, 3288 (1985).

\bibitem{kro81}
E. Kr\"oner, in
{\it Continuum Theory of Defects ---
Physics of Defects}, 
Proceedings of the XXXV Less Houches Session
1980, edited by R. Balian {\it et al}.
(North-Holland, Amsterdam, The Netherlands, 1981);
M. O. Katanaev and I. V. Volovich,
Ann. Phys. (N.Y.) {\bf 216}, 1 (1992);
R. A. Puntigam and H. H. Soleng,
Class. Quant. Grav. {\bf 14}, 1129 (1997);
F. Moraes, Braz. J. Phys. {\bf 30}, 304 (2000).

\bibitem{maz86}
P. O. Mazur, Phys. Rev. Lett. {\bf 57}, 929 (1986).

\bibitem{gal93} D. V. Gal'tsov and P. S. Letelier,
Phys. Rev. D {\bf 47}, 4273 (1993).

\bibitem{tod94} K. P. Tod,
Class. Quantum  Grav. {\bf 11}, 1331 (1994);
D. A. Konkowski and T. M. Helliwell,
Gen. Relativ. Gravit. {\bf 38}, 1069 (2006);
S. Krasnikov, Phys. Rev. D {\bf 76}, 024010 (2007).

\bibitem{des84}
S. Deser, R. Jackiw, and G. 't Hooft,
Ann. Phys. (N.Y.) {\bf 152}, 220 (1984);
G. Cl\'{e}ment, Int. J. Theor. Phys. {\bf 24}, 267 (1985).

\bibitem{mat02}
M. Visser, arXiv:gr-qc/0204022.

\bibitem{haw92}
S. W. Hawking,
Phys. Rev. D {\bf 46}, 603 (1992).

\bibitem{lor04}
V. A. De Lorenci and E. S. Moreira, Jr.,
Phys. Rev. D  {\bf 70}, 047502 (2004).

\bibitem{lor03}
V. A. De Lorenci and E. S. Moreira, Jr.,
Phys. Rev. D {\bf 67}, 124002 (2003).

\bibitem{lor05}
V. A. De Lorenci, R. Klippert and E. S. Moreira, Jr.,
Phys. Rev. D {\bf 71}, 024005 (2005).

\bibitem{mat90}
G. E. A. Matsas,
Phys. Rev. D {\bf 42}, 2927 (1990);
B. Jensen and J. Ku\u{c}era, J. Math. Phys. {\bf 34}, 4975 (1993);
V. A. De Lorenci, R. D. M. De Paola and N. F. Svaiter,
Class. Quantum Grav. {\bf 16}, 3047 (1999);
V. A. De Lorenci and E. S. Moreira, Jr.,
Phys. Rev. D {\bf 65}, 107503 (2002).

\bibitem{lor01}
V. A. De Lorenci and E. S. Moreira, Jr.,
Phys. Rev. D  {\bf 63}, 027501 (2000).

\bibitem{ger89} P. S. Gerbert and R. Jackiw,
Commun. Math. Phys.  {\bf 124}, 229 (1989).

\bibitem{chm94} P. Chmielowski,
Class. Quantum  Grav. {\bf 11}, 41 (1994).

\bibitem{ste03}
H. Stephani, D. Kramer, M. A. H. MacCallum, C. Hoenselaers and E. Herlt,
{\em Exact Solutions of Einstein's Field Equations}
(Cambridge University Press, Cambridge, England, 2003).

\bibitem{whe73}
C. W. Misner, K. S. Thorne and J. A. Wheeler,
{\it Gravitation}, 
(W. H. Freeman and Company, New York, 1973).


\bibitem{lan75}
L. D. Landau and E. M. Lifshitz, 
{\it The Classical Theory of Fields}
(Pergamon Press, Oxford, England, 1975).



\end{thebibliography}
\end{document}